\documentclass[singlecolumn, letterpaper, 10pt, prd, aps]{revtex4-1}
\usepackage[english]{babel}
\usepackage{graphicx}
\usepackage{textcomp}
\usepackage{amsmath}
\usepackage{amssymb}
\usepackage{mathrsfs}
\usepackage{hyperref}
\usepackage{tensind}
\usepackage{xcolor}
\tensordelimiter{?}
\def \ee{\end{equation}}
\def \be{\begin{equation}}
\def \eea{\end{eqnarray}}
\def \bea{\begin{eqnarray}}

\begin{document}

\title{
Collisions of Spinning Massive Particles in a Schwarzschild Background}
\author{Crist\'obal Armaza, M\'aximo Ba\~nados, \& Benjamin Koch}
\affiliation{Instituto de F\'isica, Pontificia Universidad Cat\'olica de Chile, Av. Vicu\~na Mackenna 4860, 782-0436 Macul, Santiago, Chile}
\date{\today}

\begin{@twocolumnfalse}
\begin{abstract} 
It is known that the center-of-mass energy of the collision of two massive particles following geodesics around a black hole presents a maximum. 
The maximum energy increases when the black hole is endowed with spin, and for a maximally rotating hole this energy blows up, offering, in principle, a unique probe of fundamental physics. 
This work extends the latter studies by considering that the colliding particles possess intrinsic angular momentum (spin), described by the Hanson-Regge-Hojman theory of spinning particles.
By analyzing planar trajectories of spinning particles around non-rotating black holes, 
a significant increase of the invariant collision energy is found. 
Radial turning points, causality constraints, and limitations of the theory are discussed. 
\end{abstract}
\end{@twocolumnfalse}

\maketitle


\section{Introduction}

Spinless particles follow geodesics. As such, the collisions of two spinless equal-mass particles accelerated by the gravitational field of a Schwarzschild black hole produce center-of-momentum energies that are bounded by the maximum value $E_\text{cm} = 2\sqrt 5\, m$, where $m$ is the mass of the particles \citep{baushev09}. On the other hand, if the black hole is endowed with spin (Kerr black hole), then $E_\text{cm}$ may reach arbitrarily high values for collisional Penrose processes \citep{piran75, piran77}. More recently, it was shown that arbitrarily high energies also appear in collisions taking place right outside the event horizon \citep{banados09}. The latter happens when any of the two colliding particles reaches the event horizon with the critical angular momentum of a prograde geodesic orbit. In both processes involving ultrahigh energies, the black hole must be extremal, i.e, $a = M$, $a$ and $M$ being the black hole spin and mass.

Although fascinating, an unbounded collision energy under the latter conditions might be physically unfeasible. As pointed out by previous authors \citep{berti09, jacobson10}, the necessity of a maximally rotating black hole precludes the occurrence of such energetic collisions, given that extremal black holes are unlikely to exist in nature. Backreaction and gravitational radiation are also factors that are expected to prevent these unbounded energies from arising \citep{berti09}. Further, it was later shown in Ref. \citep{bejger12} that even in the ideal case considered in Ref. \citep{banados09}, the maximum energy carried by a photon that actually escapes the collision is only $\sim 1.3$ times the total initial energy of the colliding particles. For more recent discussions on 
the efficiency see also \cite{Schnittman:2014zsa,Berti:2014lva,Leiderschneider:2015kwa,Ogasawara:2015umo}.
Despite of discussions of physical viability and probability, this process has triggered a very rich discussion on non-trivial trajectories and collision energies.

The idea of this paper is to investigate on the possibility whether similar acceleration processes are possible for non-rotating black holes by virtue of taking the trajectory of spinning massive particles. 

This paper is organized as follows. In Section \ref{secTop}, the equations of the spinning top are introduced. Section \ref{secSS} summarizes the solution of those equations in a static Schwarzschild background. In Section \ref{secColl}, the collision energy of two spinning tops in Schwarzschild is calculated. Section \ref{secTraject} deals with an analysis of the trajectories, showing the existence of radial turning points. In Section \ref{sec_vel}, the discussion is complemented by analyzing the appearance of timelike to spacelike transitions. Finally, the conclusions of this paper are given in Section \ref{secConcl}. Throughout this work, signature $(-,+,+,+)$ is assumed for the metric tensor $?g_\mu\nu?$, Greek indexes run from 0 to 3, Latin indexes run from 1 to 3 unless otherwise stated, and both the gravitational constant and the speed of light are set equal to unity.

\section{Spinning particles in General Relativity}\label{secTop}

The study of the motion of spinning tops (STOPs) in the framework of general relativity started with the seminal works by Mathisson \citep{mathisson37} and Papapetrou \citep{papapetrou51} (hereafter MP), and was later  taken further by Tulczyjew \citep{tulczyjew59}, Taub \citep{taub64}, and Dixon \citep{dixon64}. In the MP realization, the particle is endowed with a finite extension much shorter than the characteristic length of the spacetime, thus possessing a dipolar moment in addition to the monopolar one defining point particles. The trajectory is described by an arbitrary (but conveniently chosen) event $X^\mu$ within the body, and the equations of motion are then derived by means of a variational procedure. Noticeably, the resulting MP equations show that a spinning test particle does \emph{not} follow geodesics, a spin-gravity coupling accounting for such a deviation. This interaction is construed as the existence of tidal forces acting on the particle. 

Alternatively, the equations of motion for a STOP may be derived from a Lagrangian theory \citep{hanson74, hojman75, hojman77}. A Lagrangian formalism provides several advantages over the classical MP formalism \citep{hojman16}. Unlike the MP equations, from a Lagrangian perspective the equations of motion are naturally reparametrization-covariant. Also, a Lagrangian theory allows to properly define the momentum of the particle, instead of introducing an ad-hoc, or a posteriori definition. The latter advantage is especially useful for a STOP, for which the momentum and the velocity are not in general parallel. In addition, a Lagrangian approach provides a natural interpretation of angular velocity, from which the spin is straightforwardly interpreted as a three-dimensional rotation. In turn, the Casimir operators of the Poincar\'e group formed with the canonical momenta are, as shall be seen later, 
constants of motion, this way providing an elegant definition for mass and spin of the test particle. All these advantages make us pick the Lagrangian theory as the starting point to describe the motion of a spinning test particle.

The equations of motion derived from the Lagrangian theory for a spinning particle read \citep{hojman75, hojman77} 
\begin{align}\label{eom1}
\frac{DP^\mu}{D\tau} &= - \frac{1}{2}?R^\mu_\nu\alpha\beta?u^\nu?S^\alpha\beta?,\\
\frac{D?S^\mu\nu?}{D\tau} &= ?S^\mu\beta??\sigma_\beta^\nu? - ?\sigma^\mu\beta??S_\beta^\nu? = P^\mu u^\nu - u^\mu P^\nu,\label{eom2}
\end{align}
where $\tau$ is an affine parameter, $u^\mu = dX^\mu/d\tau$ is the tangent vector to the trajectory, $?\sigma^\mu\nu?$ is the angular velocity, $P^\mu$ is the canonical momentum, $?S^\mu\nu?$ is the canonical spin tensor, $?R^\mu_\nu\alpha\beta?$ is the Riemann tensor, and $D/D\tau \equiv u^\mu\nabla_\mu$ is the covariant derivative along $u^\mu$. The second equality in Eq. \eqref{eom2} comes from the definition of the canonical momenta in this theory \citep{hojman75}. One then defines the spin modulus as
\begin{equation}
S^2 = \frac{1}{2}?S_\mu\nu??S^\mu\nu?,
\end{equation}
and the mass of the particle as
\begin{equation} 
m^2 = -P_\mu P^\mu,
\end{equation}
and, up to this point, only $S^2$ is conserved by virtue of the equations of motion alone. 
The conservation of $S^2$ can easily be seen from contracting (\ref{eom2}) with $S_{\mu \nu}$.
Please note that in the Lagrangian formalism, this conservation of $S^2$ appears naturally,
whereas in formulations like \cite{dixon64}, an additional supposition is necessary.
Further, the quantity
\begin{equation}
Q_\xi = P^\mu\xi_\mu - \frac{1}{2}?S^\mu\nu??\xi_\mu;\nu?,
\end{equation}
where $\xi^\mu$ is a Killing vector of the spacetime, is another constant of motion regardless of the background metric. 

A quick glance at the equations of motion reveals that there are more unknowns that equations. For this, three of the six components of the spin tensor, namely $?S^0i?$, may be arbitrarily set equal to zero in one specific frame of reference. This freedom follows from the arbitrariness of how one fixes the worldline describing the trajectory of the particle (\citep{costa15} and references therein). The elements $?S^0i?$ define the mass dipole of the particle, so demanding it to be zero in one specific frame defines the center of mass in that frame. This frame is chosen to be the one moving with 4-velocity $P^\mu/m$, for which the \emph{covariant} ``spin supplementary condition'' reads 
\begin{equation}\label{tul}
?S^\mu\nu?P_\nu = 0,
\end{equation}
or $?S^0i? = 0$ in the zero 3-momentum frame. It can actually be shown that Eq. \eqref{tul}  defines a \emph{unique} worldline $X^\mu(\tau)$ \citep{beiglbock67, schattner79}. (See Ref. \citep{costa15} for a thorough review on different choices of spin supplementary conditions.) An important consequence from Eq. \eqref{tul} combined with the equations of motion \eqref{eom1} and \eqref{eom2}, is that $m^2$ is now a constant of motion as well. 
Nevertheless, the invariant $u_\mu u^\mu$ is \emph{not} in general a constant of motion.

\section{Spinning particles in Schwarzschild Background}
\label{secSS}
This work concentrates on the motion of a spinning particle in the vicinity of a static, uncharged black hole. The analysis of spinning particles moving around Schwarzschild black holes was first carried out by Corinaldesi \& Papapetrou \citep{corinaldesi51} solving the MP equations, and later by Hojman \citep{hojman75} solving the resulting equations derived from the Lagrangian formalism. An analytic treatment of the trajectories in general Schwarzschild-like spacetimes is possible \citep{zalaquett14}, although we shall not deal with them for simplicity. 

The background spacetime considered here is characterized by the line element
\begin{equation}
ds^2 = -\left(1-\frac{2M}{r}\right)dt^2 + \frac{dr^2}{1 - 2M/r} + r^2\,d\Omega^2,
\end{equation}
described in usual Schwarzschild coordinates $(t,r,\theta,\phi)$, where $d\Omega^2 = d\theta^2 + \sin^2\theta\,d\phi^2$. From the symmetries of this particular background, it follows that the energy of the particle is conserved along the motion. Also, the three (Cartesian) spatial components of the total angular momentum (that is, orbital angular momentum plus spin) are conserved. The latter implies that there exist equatorial ($\theta = \pi/2$) trajectories in which the spin remains perpendicular to the plane of motion \citep{hojman75}. For simplicity, the present paper focuses on these trajectories, for which the non-vanishing components of the momentum are \citep{hojman75, hojman13a}
\begin{align}\label{Pt}
\frac{P^t}{m} &= \left(1 - \frac{2M}{r}\right)^{-1}\frac{e - Mjs/r^3}{1 - Ms^2/r^3},\\\label{Pphi}
\frac{P^\phi}{m} &= \frac{1}{r^2}\left(\frac{j - es}{1 - Ms^2/r^3}\right),\\\label{Pr2}
\left(\frac{P^r}{m}\right)^2 &= \left(\frac{e- Mjs/r^3}{1 - Ms^2/r^3}\right)^2 -\left(1 - \frac{2M}{r}\right)\left[1 + \frac{1}{r^2}\left(\frac{j- es}{1 - Ms^2/r^3}\right)^2\right],
\end{align}
where $e = E/m$, $j = J/m$ and $s = \pm S/m$ are the specific energy, total angular momentum and spin
per unit mass, respectively. $s$ positive (negative) means that the spin is (anti)parallel to the total angular momentum. Analogous expressions for the non-trivial components of the spin in terms of the coordinate radius $r$ follow from this formalism. It is found that the spatial component of the spin perpendicular to the plane of motion is
\begin{equation}
S_z = r?S^r\phi? = ms\left(\frac{e - Mjs/r^3}{1 - Ms^2/r^3}\right),
\end{equation}
such that $S_z/m = se$ is the spatial component of the specific spin far away from the black hole ($M/r\longrightarrow 0$). Finally, the coordinate velocities are found to be
\begin{align}\label{urp}
\frac{dr}{dt} &\equiv \frac{u^r}{u^t} = \frac{P^r}{P^t},\\ \label{urf}
\frac{d\phi}{dt} &\equiv \frac{u^\phi}{u^t} = \left(\frac{1 + 2Ms^2/r^3}{1 - Ms^2/r^3}\right)\frac{P^\phi}{P^t}.
\end{align}
As usual, the parameter $\tau$ needs to be fixed by an external condition in order to obtain $u^t$, $u^r$, and $u^\phi$. However, for the presented discussion of relativistic invariants no such choice is needed.

\section{Collision Energy}\label{secColl}

Spinless, equal-mass particles are known to have a maximum collision energy \citep{baushev09}. 
This maximum increases when the hole rotates, diverging when the hole is maximal \citep{banados09}. It is the purpose of this paper to investigate the effect of the spin of the particles on the energy. As a function of the coordinate radius, the energy of the collision of two spinning particles 1 and 2 falling from infinity at rest is
\begin{equation}
E_\text{cm}^2 = -(\vec P_1 + \vec P_2)^2 =  m_1^2 + m_2^2 - 2\,\vec P_1 \cdot \vec P_2. \label{Ecm}
\end{equation}
By direct replacement of Eqs. (\ref{Pt}-\ref{Pr2}) in Eq. \eqref{Ecm}, the collision energy of two particles with masses $m_1 = m_2 = m$ reads
\begin{widetext}
\begin{align}\nonumber
E_\text{cm}^2 &= \frac{2m^2}{\Delta_1\Delta_2\Delta}\Bigg\{r(r^3-Mj_1s_1)(r^3-Mj_2s_2) + \Delta\left[\Delta_1\Delta_2 - r^4(j_1-s_1)(j_2-s_2)\right]\\
&\qquad\qquad\qquad-\sqrt{r(r^3-Mj_1s_1)^2-\Delta[\Delta_1^2 + r^4(j_1-s_1)^2]}\sqrt{r(r^3-Mj_2s_2)^2-\Delta[\Delta_2^2+r^4(j_2-s_2)^2]}\Bigg\},\label{Ecm_sch}
\end{align}
\end{widetext}
where $\Delta \equiv r - 2M$ and $\Delta_i \equiv r^3 - M s_i^2$, $i = 1$, 2. From this expression, it is seen that $E_\text{cm}$ could possibly diverge at the Schwarzschild radius, $\Delta = 0$, as well as at a spin-related radius, $\Delta_i = 0$. For the former, the divergent terms are canceled by vanishing terms in the numerator, and the energy in this limit is
\begin{widetext}
\begin{equation}
\lim_{r\rightarrow 2M}\frac{E^2_\text{cm}}{m^2} = \frac{\left[(8M^2 - j_1 s_1)\Delta_2 + (8M^2 - j_2 s_2)\Delta_1\right]^2+ 16M^2\left[(j_1 - s_1)(8M^2 - j_2 s_2) - (j_2 - s_2)(8M^2 - j_1 s_1) \right]^2}{\Delta_1\Delta_2(8M^2 - j_1s_1)(8M^2 - j_2 s_2)},\label{Ecm_hor}
\end{equation}
\end{widetext}
which reduces to the known case  $E_\text{cm}(r\rightarrow 2M) = (m/2)\sqrt{16 + (j_1 - j_2)^2/M^2}$ 
\citep{baushev09} when $s_i = 0$ (only for Eq. \eqref{Ecm_hor}, $\Delta_i = 8M^2 -s_i^2$). 
The case $\Delta_i = r^3 - Ms_i^2 = 0$ is more interesting because one can show that for arbitrary values of $j_i$ and $s_i$, no such cancellation takes place. 

In section \ref{sec_vel} it will be shown that this divergence occurs only 
after at least one of the trajectories passes from timelike to spacelike. This fact of potentially divergent invariant quantities, such as the collision energy, is on its own an interesting observation. However, it might be that those divergences are either not reached by physical trajectories, or that they are not observable because they occur behind the event horizon. This question will be discussed in the following section.

\section{Characterization of radial turning points}\label{secTraject}

\begin{center}
\begin{figure}
\begin{minipage}{13.2cm}
\includegraphics[width=6.5cm]{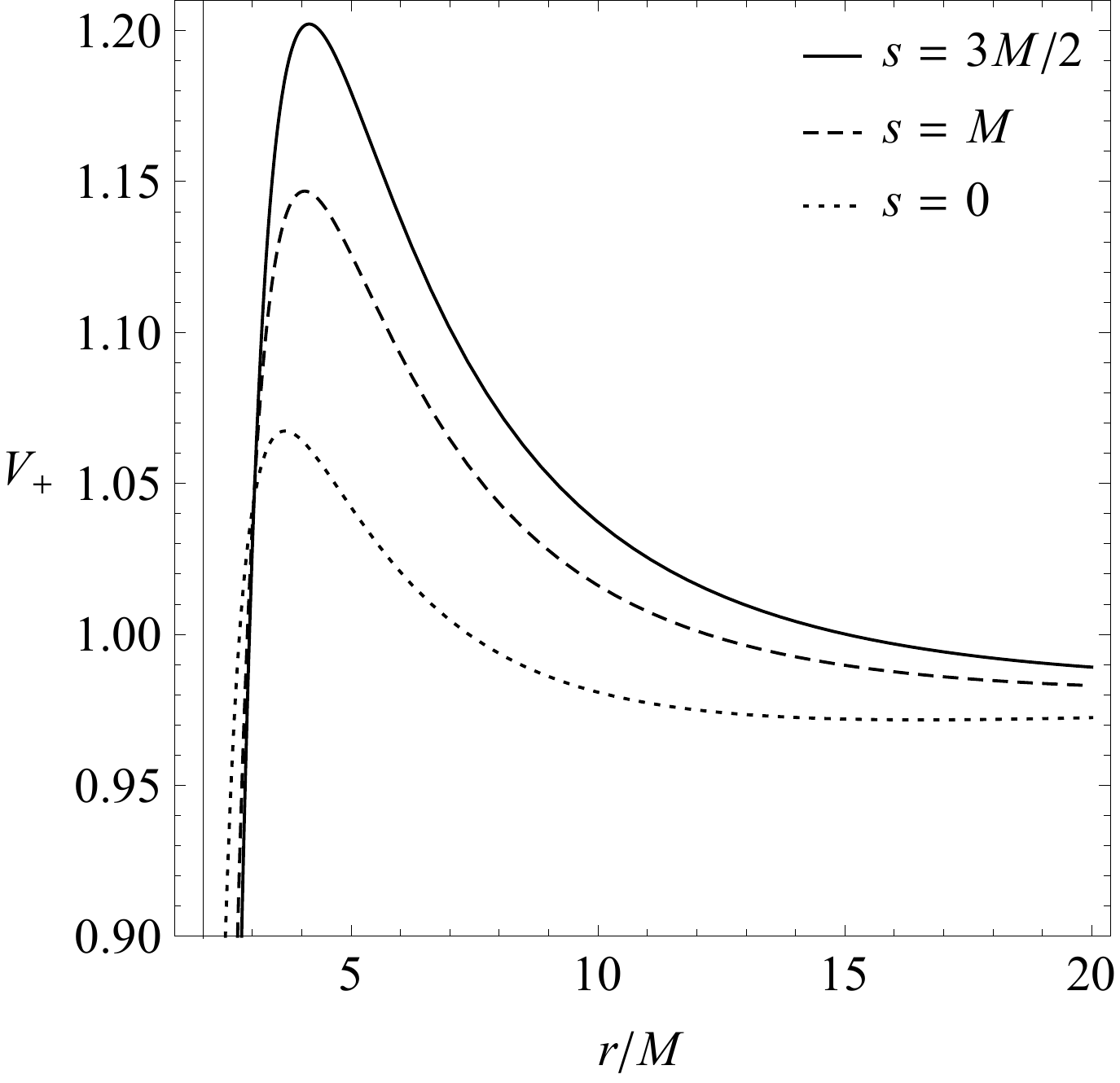}\hspace{0.2cm}\includegraphics[width=6.5cm]{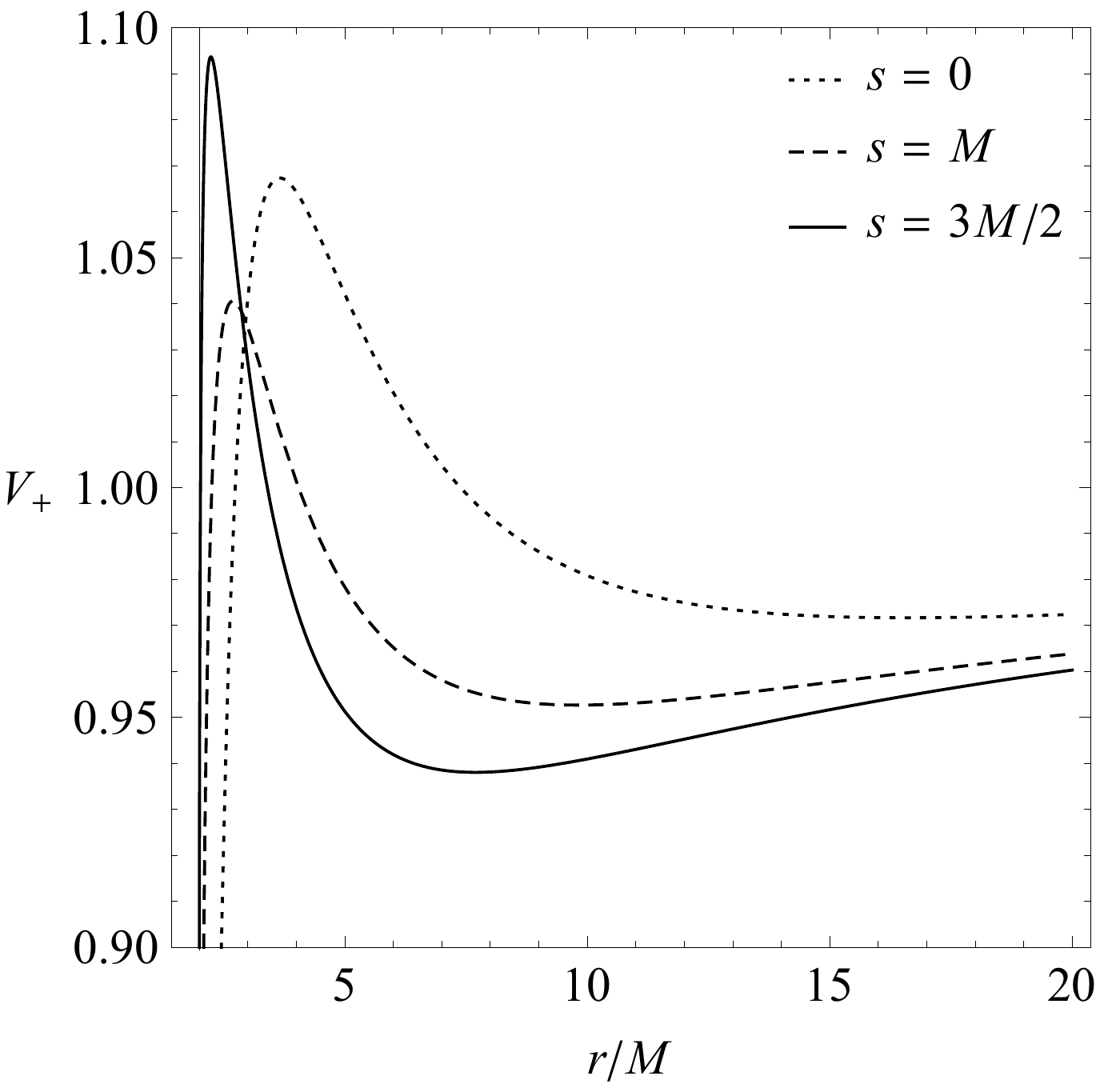}
\caption{
Effective potential $V_+(r)$ for given $j$ and different values of spin. Left: trajectories with $j = -4.5M$. Right: trajectories with $j = 4.5M$. The vertical, dotted line in each plot marks the event horizon.}\label{Vplus}
\end{minipage}
\end{figure}
\end{center}

The aim of this section is to discuss the question whether a physical trajectory actually reaches the divergence radius
\begin{equation}\label{rsing}
r_s \equiv M\left(\frac{s}{M}\right)^{2/3},
\end{equation}
or shows a radial turning point before. First of all, notice that the simultaneous transformation $j\rightarrow -j$ and $s \rightarrow -s$ is equivalent to flip over the coordinate system, so without loss of generality one can only consider the case $s \geq 0$. Now, since $u^\phi \propto P^\phi$, one can see from Eq. \eqref{Pphi} that a trajectory is either ``direct'' if $l \equiv j - es > 0$ or ``retrograde'' if $l < 0$, as long as $r^3 > Ms^2$. Further, since $j$ is the specific total angular momentum everywhere at any time, and since $se$ is the spatial component of the specific spin far away from the black hole, 
$l$ can be interpreted as the orbital angular momentum far away from the black hole. 
The radial turning points can be found by requiring that $u^r = 0$. This condition introduces an ``effective potential'',  
defined as the value of the energy when $u^r = 0$. Since $u^r$ and $P^r$ are proportional, the zeroes of $P^r$ coincide with the turning points.  
Factorizing the energy dependence from the radial dependent part of $(P^r)^2$, one gets
\begin{equation}
\left(\frac{P^r}{m}\right)^2 = a\left(1 - \frac{Ms^2}{r^3}\right)^{-2}\left[e - V_+(r)\right]\left[e - V_-(r)\right],
\end{equation}
where the effective potential
\begin{equation}
V_\pm(r) = \frac{b \pm \Sigma^{1/2}}{a}
\end{equation}
was defined, with
\begin{equation}
a = 1 - \left(1 - \frac{2M}{r}\right)\frac{s^2}{r^2},\qquad\qquad
b = -\frac{js}{r^2}\left(1 - \frac{3M}{r}\right),
\end{equation}
and
\begin{equation}
\Sigma = \left(1-\frac{2M}{r}\right)\left(1-\frac{Ms^2}{r^3}\right)^2\left[1 + \frac{j^2}{r^2} - \left(1-\frac{2M}{r}\right)\frac{s^2}{r^2}\right].
\end{equation}
In order for $P^r$ (and thus $u^r$) to be real, the motion of a spinning particle is restricted to those values of $r$ such that $e> V_+(r)$ or $e<V_-(r)$ whenever $a > 0$. It is not difficult to prove that $a = 1 - s^2/27M^2$ at $r = 3M$ is the minimum value of $a$ over $r\in [2M,\infty)$, so $a$ is everywhere positive for $s^2 < 27M^2$, while $a < 0$ for some values of $r$ in the opposite case. 
 For $s=0$, one has $a=1$, $b=0$, so $V_+(r) = -V_-(r)$, which recovers the effective potential of a spinless particle. Figure \ref{Vplus} shows the effective potential for direct and retrogade trajectories, for different values of $s$. 

\begin{center}
\begin{figure}
\begin{minipage}{14.8cm}
\includegraphics[width=7cm]{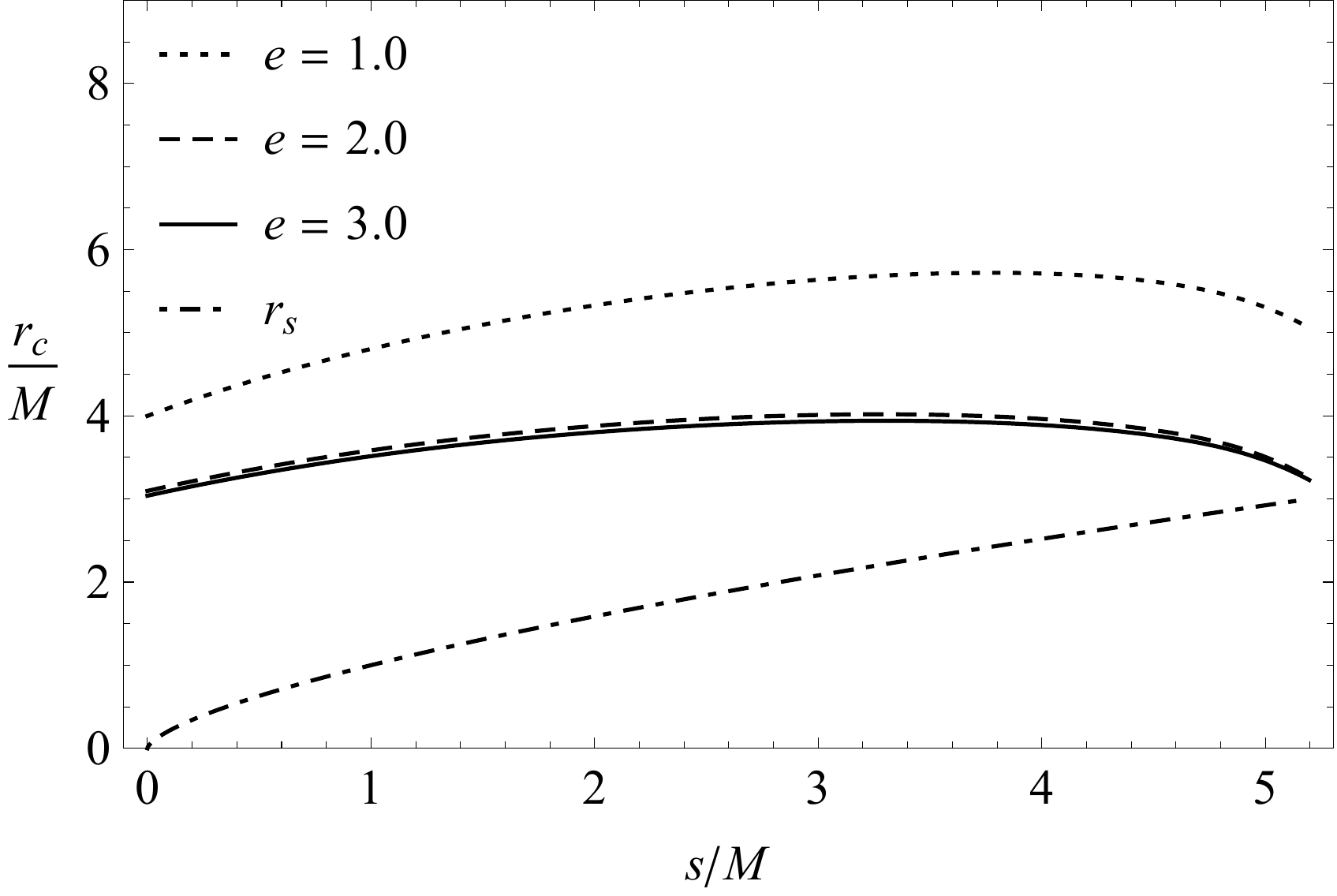}\hspace{0.8cm}\includegraphics[width=7cm]{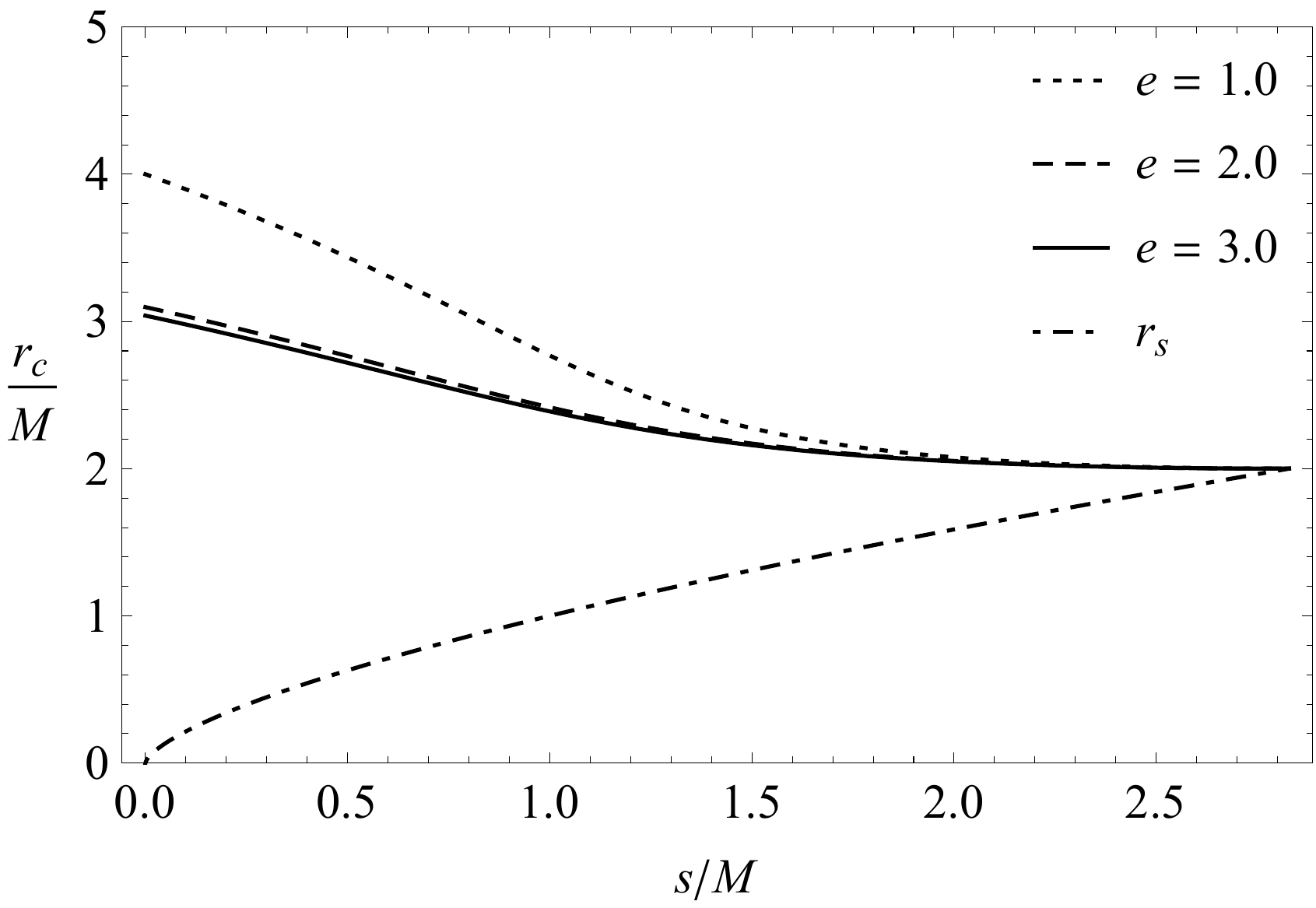}
\caption{Critical radius as a function of spin $s$ in comparison to the divergence radius $r_s$ for different energies. Left: retrograde orbits. Right: direct orbits.}\label{fig_rc}
\end{minipage}
\end{figure}
\end{center}

\begin{center}
\begin{figure}
\includegraphics[width=8cm]{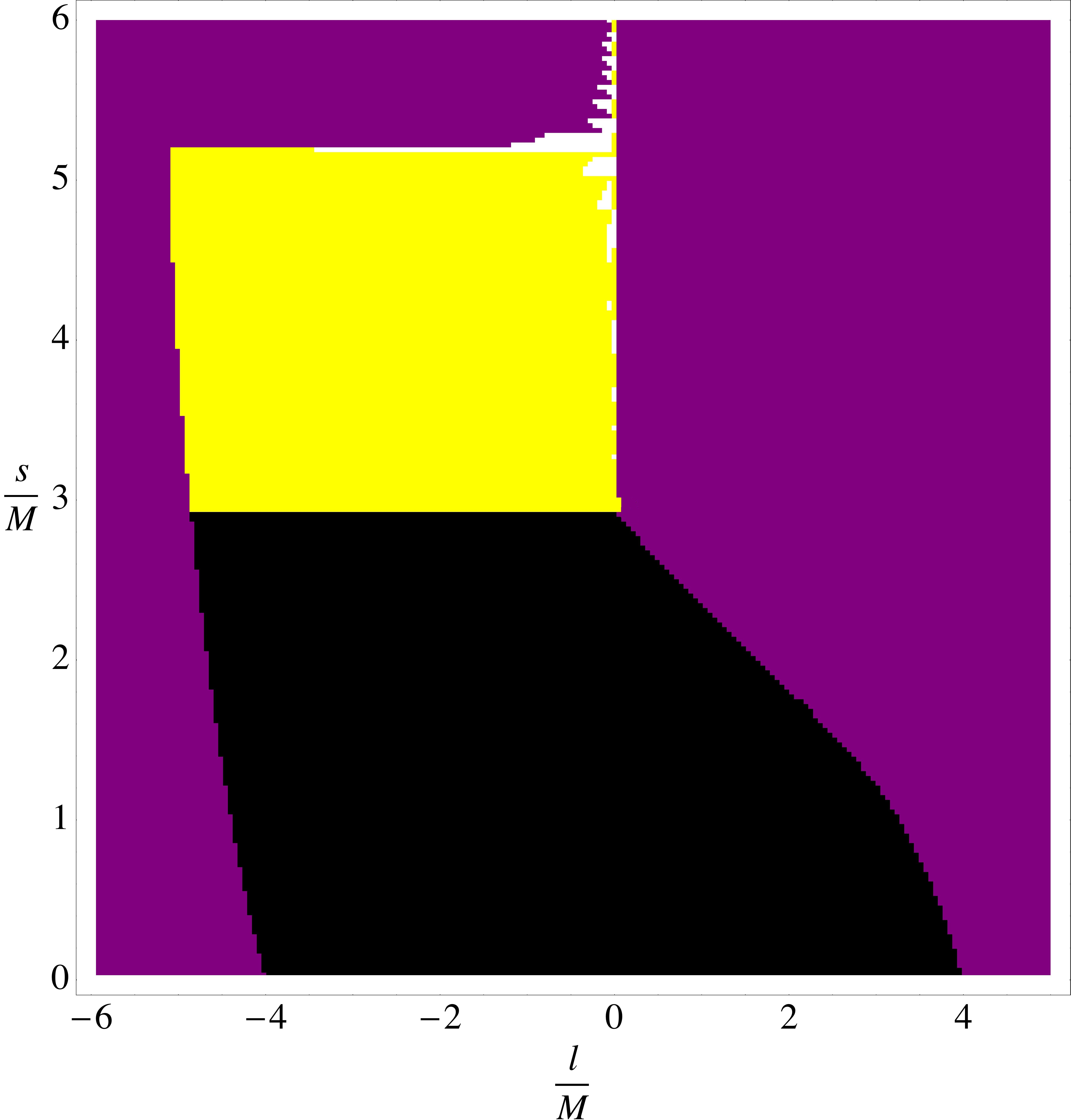}
\caption{
Characterization of the trajectories as a function of the orbital angular momentum $l$ and the spin $s$, for $e =1=M$. 
The trajectories are run inward from large $r$ to small $r$ and it is tested for each step
whether, a radial turning point, a horizon crossing, or a divergent center of mass energy (divergence radius) occurs.
The color coding is:\\
 \textcolor{purple}{purple} $=$ turning point first, 
{\bf{black}} $=$ horizon first, 
\textcolor{yellow}{yellow} $=$ divergence radius first.
White points correspond to trajectories which have the radial turning point and the divergence radius
too close together to be distinguished by the chosen numerical step size ($\delta r/r=32.000$).}\label{fig_cool1}
\end{figure}
\end{center}

For particles with positive energy, the situation is qualitatively equivalent to the spinless case: given an energy, a particle coming from infinity may either reach the horizon or meet a turning point, depending on its orbital angular momentum. As is well known, a spinless particle falling from infinity at rest ($e = 1$) with $|l| > 4M$ meets a turning point at some $r > 4M$, while it falls into the hole if $|l| < 4M$. 
The case $|l| = |l_c| = 4M$ is a critical situation where particles start to fall, one calls $r_c \equiv 4M$ the critical radius. 
If one now allows the particles to have spin, direct and retrograde orbits present asymmetric characteristics. 
\begin{itemize}
\item
For direct trajectories $(l > 0)$, an increasing spin allows a smaller $l_c$, resulting in a turning point closer to the hole (Fig. \ref{fig_rc}). As $l_c$ decreases, the critical radius $r_c$ tends (without touching) to the event horizon $r = 2M$.  For $s < 2\sqrt 2 M$, Fig. \ref{fig_rc} shows that in all cases the divergence radius $r_s$ is always inside the horizon, and thus $r_c > r_s$. Of course, $s = 2\sqrt 2M$ corresponds to the spin for which the divergence radius coincides with the event horizon. For $s > 2\sqrt 2 M$, all direct trajectories present turning points regardless of its orbital angular momentum. The divergence radius now lies outside the horizon, but none of these trajectories reaches this radius. Fig \ref{fig_cool1} presents in detail which trajectories fall into the horizon, depending on their parameters $(s, l)$, for $e = 1$.
\item
Particles following retrograde orbits $(l < 0)$ present more interesting features. As the spin increases, one sees that $|l_c|$ also increases, 
so the turning points now lie further away from the hole. 
For particles with $s < 2\sqrt 2 M$, the critical radius is a monotonic increasing function of the spin, as Fig. \ref{fig_rc} shows. In the same range of spin, orbits with $|l| \geq |l_c|$ reach the event horizon just as in the direct case. However, for trajectories with $s > 2\sqrt 2 M$, new phenomena occur. In that case, the divergence radius lies outside the event horizon, and particles with $|l_c| < l$ can actually reach this radius \emph{before reaching the horizon}. 
This happens at least in the range $ 8M^2 < s^2 < 27M^2$, for which the effective potential is clearly a real number.
 In the latter range of spin, the critical radius reaches a maximum and then decreases, as can be seen in Fig.~\ref{fig_rc}. 
The numerical characterization of the retrograde trajectories is shown on the left hand side of Fig. \ref{fig_cool1}.
\end{itemize}

As already mentioned, the center-of-mass energy of two STOPs increases without bounds at the divergence radius. Now it is clear that some \emph{retrograde} orbits may, in principle, reach the divergence radius and thus attain an infinite $E_\text{cm}$. It is worth noticing that, at this same radius, both the momenta and the $z-$component of the spin diverge as well. The appearance of divergent quantities within a theory can be a  sign of a new discovery but it can also be a sign of a possible breakdown of the applicability of the theory. 
Caution dictates to take the latter attitude. In this sense, the region in the ($l$, $s$) 
parameter space where divergent invariant quantities become imminent should be treated with care and analyzed from a different perspective.
Another perspective on the STOP trajectories can be gained by evaluating their velocities instead of their canonical momenta.
This is done in the following section.

\section{Velocity and proper time}
\label{sec_vel}
\begin{center}
\begin{figure}
\includegraphics[width=8cm]{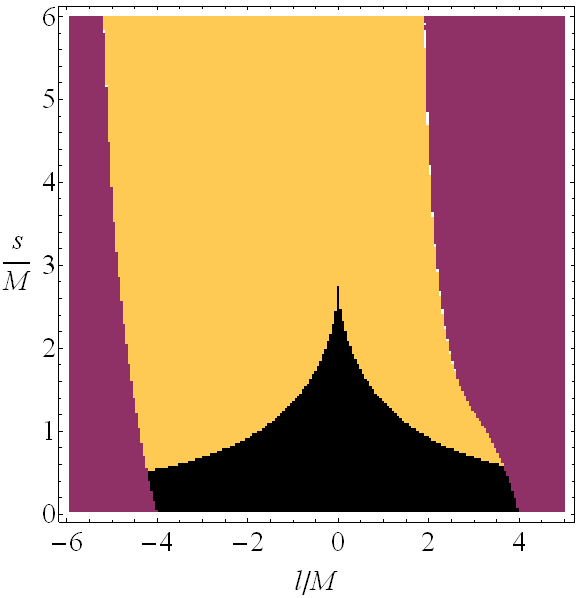}
\caption{
Characterization of the trajectories as a function of the orbital angular momentum $l$ and the spin $s$, for $e = 1=M$. 
The trajectories are run inward from large $r$ to small $r$ and it is tested for each step
whether, a radial turning point, a horizon crossing, a divergent center of mass energy (divergence radius), or a transition from
timelike to spacelike occurs.
The color coding is:\\
 $\left\{\right.$\textcolor{purple}{purple} $=$ turning point first, 
{\bf{black}} $=$ horizon first, 
\textcolor{yellow}{yellow} $=$  superluminal first$\left. \right\}$.
White points correspond to trajectories which have the radial turning point and the radius of superluminal motion
too close together to be distinguished by the chosen numerical step size ($\delta r/r=5000$).
}\label{fig_cool2}
\end{figure}
\end{center}

In the described theory of STOPs, the velocity square $u_\mu u^\mu$ is not a conserved quantity, in contrast to the momentum squared $P_\mu P^\mu=-m^2$. Therefore, the velocity square will be examined in the light of the findings for the invariant collision energy, paying close attention to its behavior in the regime where divergences appear. This invariant reads
\begin{equation}
\frac{u_\mu u^\mu}{(u^t)^2} = -\left(1 - \frac{2M}{r}\right)^2\left(\frac{1-Ms^2/r^3}{e-Mjs/r^3}\right)^2\left[1 - \frac{3Ms^2(j-es)^2}{r^5}\frac{(2 + Ms^2/r^3)}{(1-Ms^2/r^3)^4}\right].\label{u2}
\end{equation}
A particle starting its trajectory at rest ($E=m$) at radial infinity will thus have $u_\mu u^\mu= -1$. Throughout the trajectory, the velocities $u^r$ and $u^\phi$ increase with decreasing radial coordinate $r$. In particular, for certain values of $l$ and $s$ (see Fig. \ref{fig_cool2} and Eq. (\ref{u2})), $u^\phi$ will eventually grow without bounds, which would correspond to an arbitrarily fast rotation and to $u_\mu u^\mu\rightarrow + \infty$ at the divergence. However, before this happens, the trajectory will necessarily pass from timelike to spacelike. 
Thus, any trajectory that would lead to divergent collision energy would first pass from timelike to spacelike. 
The radius where the transition from timelike to spacelike occurs is of course found by solving $u_\mu u^\mu=0$. This corresponds to solve a polynomial equation of order 12, with no obvious analytical solution. 

The fact that those trajectories which have divergencies are actually trajectories
which also have a timelike to spacelike transition, is in agreement with
general studies of the underlying equations \cite{Deriglazov:2015zta}, which predict
divergencies in the limit of $v\rightarrow c$ (see also \cite{Deriglazov:2015kba}). 

From comparing Figs. \ref{fig_cool1} and \ref{fig_cool2}, one sees
that the $(l,s)$ parameter space for the superluminal-first trajectories
includes all configurations with diverging collision energy, some of the configurations
with radial turning points and some of the configurations which cross the event horizon.
In this context it would be interesting
to investigate how is the subset of trajectories (\ref{Pt}-\ref{urf}) that has
a timelike to spacelike transition without producing a diverging center of mass energy
related to the ideas of singularity avoidance of \cite{Deriglazov:2015zta}.
One further observes that not only retrograde trajectories, but also some direct trajectories
with sufficiently large spin, can have those transitions.
The velocity components $u^t,\; u^r, \, u^\phi$ further allow to calculate the proper time
of the STOP along its path, which is (assuming the usual relation)
\be
\Delta \tau= \int \sqrt{|g_{\mu \nu}dx^\mu dx^\nu|}
=\int d\tau \sqrt{\left| u^\mu u_\mu \right|}.
\ee
Inserting (\ref{u2}) into this expression 
one sees that the differential under the integral diverges with power $-1$ as $r$ approaches $r_s$
and thus, for finite $u^t$, the proper time diverges logarithmically in this limit.
This observation could be used as an argument that the STOP actually
never really reaches the divergence. Similar arguments have been used
in the context of non spinning particles in Kerr background \cite{Harada:2014vka}.

\section{Conclusions}\label{secConcl}

In this paper, the motion of STOPs in the equatorial plane of a Schwarzschild black hole is analyzed.
In particular, it is found that retrograde trajectories 
(spin angular momentum pointing in opposite direction of orbital angular momentum) 
can experience significant accelerations.
This acceleration seems to provoke divergent center-of-mass energies if the STOP collides
with another particle moving in the same plane.
This is in analogy to the similar case of a point particle being accelerated by a rotating black hole.
However, the STOPs trajectories in a Schwarzschild background seem to have further exotic features which are not present for
the point particle in a Kerr background.
The most important ones are:
first, the lighter the STOP in relation to its spin (essentially the ratio $(Sc)/(GMm)$), the larger the deviation from spinless geodesics. 
Second, there exists a continuous range of spin angular momentum $s$ and orbital angular momentum $l$ for
which divergent center-of-mass energies can occur outside of the black hole horizon.
Third, a necessary but not sufficient condition for such a divergence is that the STOP has to
pass from timelike to spacelike motion.
Fourth, the proper time of the STOP to reach $r_s$ is infinite.
In any case, it is advisable to treat the trajectories with superluminal
parts with care, since they might lie in a regime where the STOP theory ceases
to be a low energy limit of a underlying field theoretical description.

The appearance of superluminal motion and the corresponding causality violation
is a severe problem. 
One might speculate that this is an artificial effect which indicates the breakdown
of the dipolar approximation, which was used in the Lagrangian approach.
However, even if one would incorporate arbitrarily high orders in the formalism,
it is likely that similar non-causalities continue to appear.
The reason for this conjecture is that almost all field theories
have non-causal solutions when coupled to general 
relativity~\cite{Camanho:2014apa}.
Since a STOP description should be understood as an approximation to
an underlying field theory, it would be strange that going beyond second
order in the approximation would solve a problem which is actually present in the
underlying theory.

The authors suggest to interpret the results in the following way.
The effect of spin-induced acceleration, leading to moderately increased center-of-mass energies, can be taken as a solid result.
Taking this result to its extremes will most likely not be physical, since
the STOP theory should represent some low-energy limit of an underlying quantum
field theory describing the spinning particle as fluctuations of a field.
This limit is most likely to break down before superluminal motion (and thus infinite collision energies) occur.

\section*{Acknowledgments}

We thank Nicol\'as Zalaquett and Sergio A. Hojman for fruitfully spinning discussions and the referees
for their comments.
The work of B.K. and C.A. was supported by Fondecyt Project 1120360 and ANILLO ACT-1102. 
The work of M.B. was supported by Fondecyt  1141221 and ANILLO ACT-1102.

\appendix

\end{document}